# Keratoconus Recognition Using A Parameter Set Determined from IOP-Matched Scenario

Dan Lin,[a] BS, Lei Tian,[b] PhD, Shu Zhang,[a] BS, Like Wang,[a] PhD and Yongjin Zhou [a,*],PhD

[a]Shenzhen University, Health Science Center, School of Biomedical Engineering, Xueyuan Avenue 1066, Shenzhen 518055, China

[b]Beijing Institute of Ophthalmology, Beijing Tongren Eye Center, Beijing Tongren Hospital, Capital Medical University;

Beijing Ophthalmology & Visual Sciences Key Laboratory, National Engineering Research Center for Ophthalmology, Beijing 100730, China

*Address all correspondence to: Yongjin Zhou; E-mail: yjzhou@szu.edu.cn; telephone: 0755-26909971; mailing address: Shenzhen University, Health Science Center, School of Biomedical Engineering, Xueyuan Avenue 1066, Shenzhen 518055, China.

Acknowledgements: This research was supported by the National Key R&D Program of China (2016YFC0104700); the Science and Technology Planning Project of Guangdong Province (2015A020214022, 2015B020214007); National Natural Science Foundation of China (31600758, 81471735, 61427806); Beijing Natural Science Foundation (7174287); Beijing Nova Program (Z181100006218099); Beijing Municipal Administration of Hospitals' Youth Programme (QMS20170204).

Dan Lin and Lei Tian contributed equally to this work and should be considered as equal first authors.

## Abstract

**PURPOSE:** Among the current practices for keratoconus recognition using biomechanical parameters from corneal visualization Scheimpflug technology (Corvis ST), matching intra-ocular pressure (IOP) is often required to eliminate the biasing influence; as a result, the combined biomechanical parameters determined from IOP-unmatched scenario possibly bring in confounding influence. This paper was therefore designed to introduce a novel compatible parameter set (CPS) determined from IOP-matched scenario, hopefully could show its compatibility and superiority for recognizing keratoconus in both IOP-matched and not scenarios.

**METHODS:** A total of 335 eyes were included. Among them, 70 eyes (35 keratoconus and 35 normal eyes; pairwise matching for IOP) were used to determined CPS by forward logistics regression, 62 eyes (31 keratoconus and 31 normal eyes; pairwise matching for IOP) were used to validate CPS in IOP-matched scenario, and resting 203 eyes (112 keratoconus and 91 normal eyes; not pair matching for IOP) were used to validate CPS in IOP-unmatched scenario. To analyze its superiority, CPS was also compared with other two reported Biomechanical Indexes (aCBI [6,7] and DCR [8]) in both scenarios. Receiver operating characteristic curves (ROC), accuracy, FI, sensitivity and specificity were used to access and compare the performance of these three parameter sets in both scenarios.

**RESULTS:** The resulting CPS was comprised of only 3 biomechanical parameters: DA Ratio Max 1mm (DRM1), the first applanation time (AT1) and an energy loading parameter (Eload). In the IOP-matched validation, the area under ROC (AUC) reached 95.73%, with an accuracy of 95.2%, sensitivity of 93.5% and specificity of 96.8% (leave one out cross-validation). All these indicators reached 96.54%, 95.1%, 95.6% and 94.6% respectively, in the IOP-unmatched validation (leave one out cross-validation). Surprisingly, CPS performed better than other two parameter sets on a whole.

**CONCLUSIONS:** The parameter set determined from IOP-matched scenario indeed exhibit its superiority for differentiation of keratoconus and normal corneas, regardless of IOP-matched or not.

## Introduction

The detection of keratoconus remains a clinical challenge, especially the early detection. It's well documented that changes in biomechanical properties in keratoconus are postulated to occur before the disease becomes tomographically apparent due to abnormalities in stromal collagen [1,15-17]. Therefore, it has become popular to detect keratoconus with biomechanical parameters from Corneal Visualization Scheimpflug Technology (Corvis ST, Oculus, GER).

Corvis ST is a relatively new device that can cause corneal deformation with an air puff and observe the whole deformation process in real time with an ultra-high-speed Scheimpflug camera [2,3]. From the deformation process, Corvis ST can estimate corneal biomechanical parameters as well as IOP. More recently, IOP has been gradually regarded as the biasing factor [4,5] and suggested to be excluded to enable unbiased analysis. Unfortunately, there is no study to determine the combined biomechanical parameters from the perspective of IOP matching to analyze its discriminative ability currently. Meanwhile, previously reported parameter sets [6-8] missed several new and important biomechanical parameters [2,9].

Therefore, in this study, we seek to determine a new parameter set combines new biomechanical parameters, based on IOP-matched scenario. The hope is the parameter set has high accuracy for keratoconus recognition in IOP-matched as well as unmatched scenarios, and subsequently further improve current keratoconus detection practices.

## Patients and Methods

*Patients*

This study included a total of 335 eyes, including 178 keratoconic eyes and 157 normal eyes. Among them, 132 eyes (66 keratoconic eyes and 66 normal eyes) were pairwise matched for IOP and 203 eyes were not matched for IOP (112 keratoconic eyes and 91 normal eyes). To enable as possible as unbiased analysis, 70 eyes after pairwise matched for IOP were randomly selected to determine parameter set. The resting 62 eye matched for IOP and 203 eyes unmatched for IOP were used to test the robustness and compatibility of the parameter set. In patients who were diagnosed with keratoconus in only one eye, that eye was selected for measurement. For participants with keratoconus in both eyes and for normal subjects, one eye was randomly selected for measurement and statistical analysis.

All patients underwent a complete ophthalmic examination, including a detailed assessment of uncorrected distance visual acuity, corrected distance visual acuity, slit-lamp microscopy and fundus examination, corneal topography (Allegro Topolyzer; Wavelight Laser Technologie AG, Erlangen, Germany), corneal tomography (Pentacam; Oculus Optikgeräte GmbH), ocular biomechanics, and IOP measurement (Corvis ST). All measurements were taken by two trained ophthalmologists during a single visit. A diagnosis of keratoconus was made if the eye had an irregular cornea determined by distorted keratometry mires or distortion of the retinoscopic or ophthalmoscopic red reflex and at least one of the following slit-lamp signs: Vogt's striae, Fleischer's ring with an arc > 2 mm, or corneal scarring consistent with keratoconus [10-12].

Potential subjects were excluded from the study if they had undergone previous corneal or ocular surgery, had any ocular pathology other than keratoconus, or had systemic diseases known to affect the eye. Participants were instructed to remove soft contact lenses at least 2 weeks and rigid contact lenses at least 1 month before the examination. Data were collected during 2017 to 2018 in the Beijing Institute of Ophthalmology, Beijing Tongren Hospital, Beijing, China. All participants signed an informed consent form in accordance with the tenets of the Declaration of Helsinki.

*Collection of Parameters*

A total of 21 biomechanical parameters were extracted from Corvis ST software and videos, including 11 parameters from Corvis ST, 9 parameters proposed by us before and 1 parameter defined by other scholars.

The Corvis ST allows noninvasive imaging of the cornea's dynamic deformation in response to an air puff. A high-speed Scheimpflug camera records the deformation with full corneal cross-sections, which are then displayed in slow motion on a control panel. During the deformation response, a precisely metered air pulse causes the cornea to move inward or flatten (the phenomena of corneal applanation), that is, the first applanation (A1). The cornea continues to move inward until reaching a point with the highest concavity. After that, it rebounds from this concavity to another point of applanation (A2), and then to its normal convex curvature. When the deformation process is complete, the Corvis ST produces several output measurements. All these parameters are explained in **Table 1** and would be assessed in the present paper.

Table 1. Parameters from the Corvis ST Software and Corresponding Definitions.

| Parameters | Abbreviation | Definitions |
|---|---|---|
| The time of the first applanation (ms) | AT1 | Time from the initiation of the air puff until the first applanation |
| The time of the second applanation (ms) | AT2 | Time from the initiation of the air puff until the second applanation |
| The length of the first applanation (mm) | AL1 | Corneal length of the first applanation |
| The length of the second applanation (mm) | AL2 | Corneal length of the second applanation |
| The velocity of the first applanation (m/s) | $V_{in}$ | Speed of corneal apex at the first applanation |
| The velocity of the second applanation (m/s) | $V_{out}$ | Speed of corneal apex at the second applanation |
| The time of highest concavity (ms) | HC-time | Time from the initiation of the air puff until the highest concavity of the cornea |
| DA Ratio Max 1 mm | DRM1 | Maximum ratios of deformation at the apex divided by the average deformation 1 mm to either side of the apex |
| Peak distance (mm) | PD | Distance of the two knees at the highest concavity |
| Deformation amplitude (mm) | DA | Maximum deformation amplitude at the highest concavity |
| The highest radius of concavity (mm) | HC-radius | Corneal concave curvature at the highest concavity |

In 2014, we proposed several new parameters to indicate the ocular biomechanical behavior [9]. Maximum area of deformation (MA) describes the maximum corneal deformation area with the two knees. Maximum area–time of deformation (MA-time) represents the time from deformation starting to the maximum deformation area occurred. Corneal contour deformation (CCD) describes a distance between the original contour and contour with the highest concavity. Maximum corneal inward/outward velocity ($V_{inmax}$/$V_{outmax}$) represents the maximum corneal inward/outward deformation velocity at centerline. Subsequently, in 2016, we proposed the Energy absorbed area ($A_{absorbed}$) and Tangent stiffness coefficient ($S_{TSC}$) to indicate the corneal viscosity and elasticity, respectively [2]. Simultaneously, the Energy loading (Eload) and Energy return (Ereturn) of cornea during the air puff indentation were defined as well, which also describe the corneal viscoelasticity.

Stiffness parameter (SP-A1) is a parameter associated with corneal stiffness [13], which is defined as resultant pressure (Pr) divided by amplitude of deformation at A1. Pr is defined as the adjusted pressure at A1 (adj-AP1) minus a biomechanically corrected IOP (bIOP) [14]. The computational equation is given by: SP-A1 = (adj-AP1 – bIOP)/ deformation amplitude at A1.

*Statistical Analysis*

Statistical analysis was performed using SPSS software (Windows 21.0) and R (Rstudio 3.4.0). The Kolmogorov–Smirnov test was used to assess a normal distribution of parameters, which are here provided as the mean and standard deviation (SD). Welch's modified Student's two-sample t-test and the Mann-Whitney U test were used to evaluate the difference of parameters between keratoconus and normal corneas. $P < 0.05$ was statistically considered significant. Forward logistic regressions were performed to determine CPS from all these biomechanical parameters in IOP-scenario. AUC, F1, sensitivity, specificity and accuracy were used to evaluate and compare the discriminative ability of parameter sets in both IOP-matched and -unmatched scenarios. The closer to 1 these indicators are, the better the parameter set is.

**Results**

CPS contained 3 parameters: DRM1, AT1 and Eload. The equation is presented as follows:

Beta = ((A1 * DRM1) + (A2 * AT1) + (A3 * Eload) +A4)

Possibility = exp (Beta) / (1+exp (Beta))

In this study, bIOP was used to correct IOP from Corvis ST based on finite element modeling [23]. As shown in **Table 2**, all parameters were significantly different between keratoconus and normal corneas ($P < 0.05$) except for bIOP and age.

The influence of single parameter within CPS is displaced in **Figure 1**. When Eload was included in logistics regression, AUC was 96.8%, with a sensitivity of 91.4% and specificity of 88.6%. After including AT1, AUC and specificity were improved to 98.0% and 91.4%, respectively. Finally, when DRM1 was included, CPS exhibited AUC, accuracy, sensitivity and specificity of 98.5%, 94.3%, 94.3% and 94.3%, respectively. (best cut-off point of 0.5) (**Figure 1**,

**Table 3**)

Table 2. Comparing the bIOP [23], CCT and Biomechanical Parameters in the Keratoconus and Normal corneas in IOP-matched Scenario (35 keratoconus and 35 normal eyes; pairwise matching for bIOP).

| Parameters | Keratoconus (n=35) | Normal (n=35) | P |
|---|---|---|---|
| bIOP [23] | 13.89±1.23 | 13.45±1.02 | 0.144 * |
| age | 23.49±7.05 | 23.66±4.21 | 0.902 # |
| CCT | 489.20±27.66 | 532.92±26.64 | 0.000 * |
| AT1 | 6.30±0.34 | 6.90±0.32 | 0.000 * |
| DRM1 | 1.18±0.02 | 1.17±0.01 | 0.012 * |
| Eload | 108.50±9.13 | 90.71±6.67 | 0.000 # |

[23] IOP from Corvis ST was corrected based on finite element modeling
* Two-tailed Student's t-test
# Mann-Whitney U test

The ROC curves of CPS, adjusted Corvis Biomechanical Index (aCBI) [7] as well as Dynamic Corneal Response Index (DCR) [8] in both IOP-matched and -unmatched validation are shown in **Figure 2**. AUC of CPS in IOP-unmatched validation was slightly higher than that in IOP-matched scenario (96.74% and 95.73%, respectively). When compared with aCBI and DCR, CPS exhibited highest AUC in both validation.

As shown in **Figure 3**, accuracy, F1, sensitivity and specificity were compared between these three parameter sets in both IOP-matched and -unmatched validation. In IOP-matched validation, CPS showed same accuracy, F1, sensitivity and specificity with aCBI (95.2%, 95.1%, 93.5% and 96.8%, respectively), while performance better than DCR. In IOP-unmatched scenario, all these four indicators of CPS were highest compared with the corresponding indicators in aCBI and DCR. Accuracy of CPS reached 95.1%, while aCBI and DCR reached 91.1%. F1 of CPS was highest at 95.5%, compared with aCBI and DCR (91.7% and 91.8%). Sensitivity and specificity of CPS also showed highest value (95.6% and 94.6%, respectively) among all three parameter sets. (cut off point is 0.5)

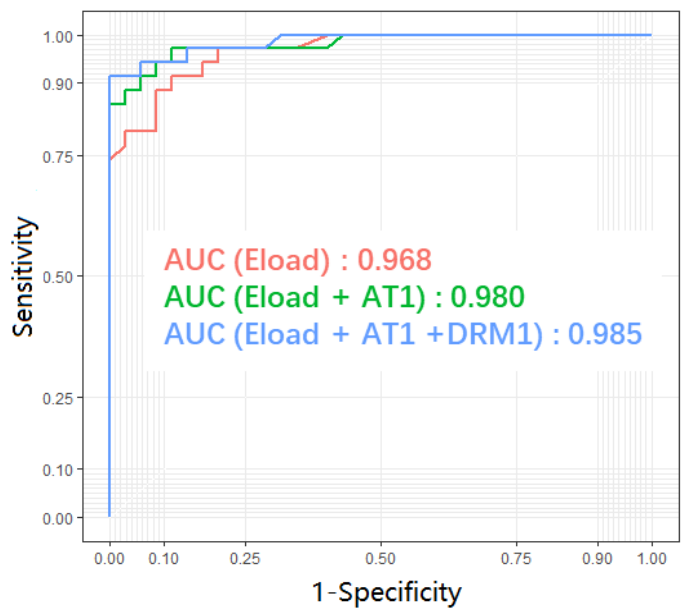

**Figure 1.** Receiver Operator Characteristic Curve for Each Step of the Forward Logistics Regression in IOP-matched Scenario (35 keratoconus and 35 normal eyes; pairwise matching for bIOP).

**Table 3.** Gain in Sensitivity and Specificity with Each Step of the Logistics Regression in IOP-matched Scenario (35 keratoconus and 35 normal eyes; pairwise matching for bIOP) (best cut-off point = 0.5).

|  |  | Observed |  |  |
|---|---|---|---|---|
|  | **Predicted** | **Keratoconus** | **Normal** | **Overall %** |
| **Step 1: Eload in** | Keratoconus | 32 | 3 |  |
|  | Normal | 2 | 31 |  |
|  | Correct % | 91.4 | 88.6 | 90.0 |
| **Step 2: AT1 in** | Keratoconus | 32 | 2 |  |
|  | Normal | 2 | 32 |  |
|  | Correct % | 91.4 | 91.4 | 91.4 |
| **Step 3: DRM1 in** | Keratoconus | 33 | 2 |  |
|  | Normal | 2 | 33 |  |
|  | Correct % | 94.3 | 94.3 | 94.3 |

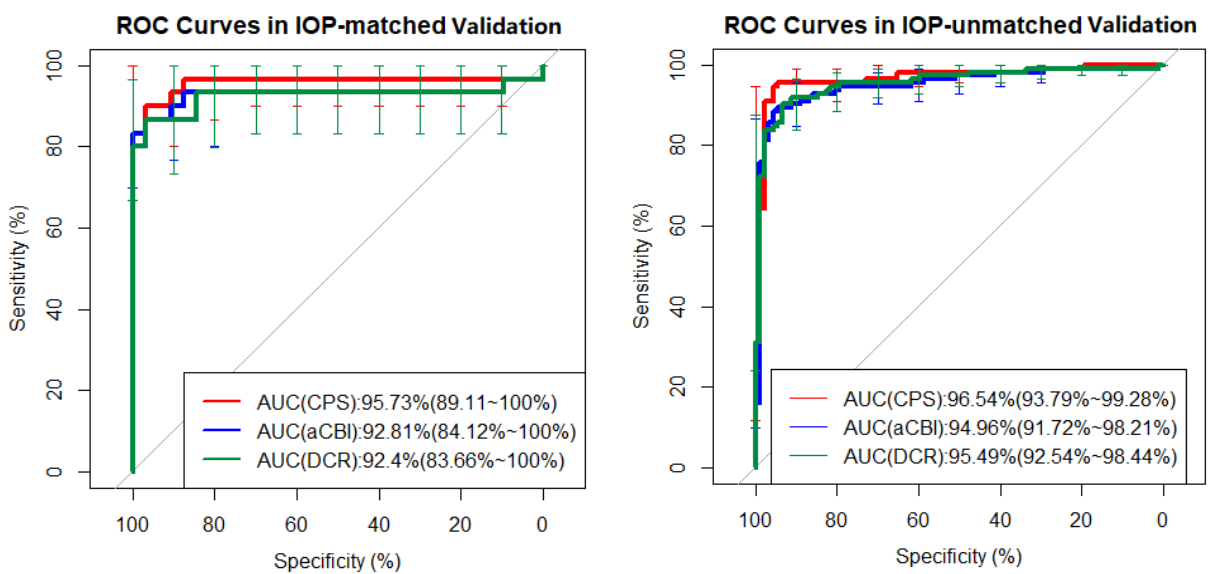

**Figure 2.** Receiver Operator Characteristic Curve for CPS, aCBI [7] and DCR [8], in IOP-matched (31 keratoconus and 31 normal eyes; pairwise matching for bIOP) and -unmatched (112 keratoconus and 91 normal eyes; not pair matching for bIOP) validation.

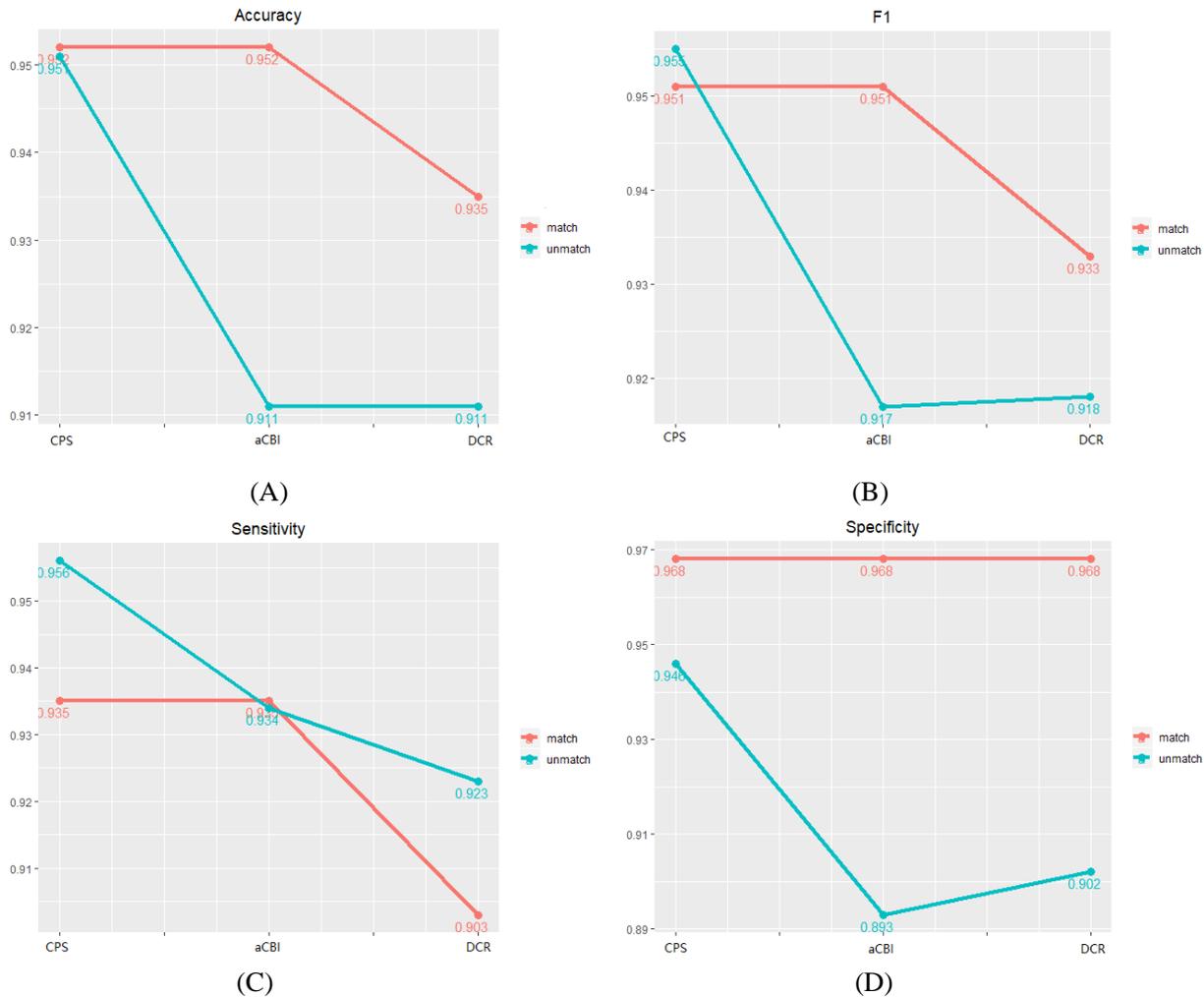

**Figure 3.** Accuracy, F1, sensitivity and specificity of CPS, aCBI [7] and DCR [8], in IOP-matched (31 keratoconus and 31 normal eyes; pairwise matching for bIOP) and -unmatched (112 keratoconus and 91 normal eyes; not pair matching for bIOP) validation.

## Discussions

To date, there are limited number of studies focused on a parameter set that could cover both scenarios of IOP-matched and -unmatched in terms of keratoconus recognition [6,7]. Vinciguerra *et al.* reported parameter set [6] established on the dataset unmatched for IOP, while validated on dataset of IOP matched [7], finding the accuracy declined. From our point of view, this decline is not only due to the matching of CCT as assumed by the authors [7], but also to the fact that the parameter set was established on IOP-unmatched dataset. In this paper, a dataset independent of IOP was deliberately used to determine a parameter set free of biasing effect of IOP, while other two datasets uninfluenced and influenced by IOP were used to validate the discriminative ability of the parameter set instead. Interestingly, the results exhibited slightly improved accuracy in the validation, which in turn confirmed the rationality of our strategy.

Meanwhile, the parameters used in previous literatures [6-8] were limited to those available from Corvis ST software. In this paper, we also incorporated more parameters from third-part reports [2,9] to enable the analysis as comprehensive as possible. As displayed in **Table 3**, only 3 parameters (DRM1, Eload, AT1) were selected in IOP-matched group finally and Eload is from a third-party report [2]. Although sensitivity remained unchanged when AT1 was included, specificity was significantly improved. It's likely that this parameter can exhibit the viscoelasticity in keratoconus well after eliminating the influence of IOP.

The ROC curve for the CPS demonstrated a slight decrease in AUC in both validation datasets (**Figure 2**) compared with training set (**Figure 1**). Yet surprisingly, at the best cut off point 0.5, accuracy was improved in both validation set (95.2% in IOP-matched validation, 95.1% in IOP-unmatched validation; while 94.3% in IOP-matched training), indicating that CPS can adapt to different datasets, regardless of IOP-matched or unmatched. What's more, as shown in **Figure 3**, CPS showed similar discriminative ability between IOP-matched and -unmatched validation (accuracy and specificity were higher in IOP-matched validation; F1 and sensitivity were higher in IOP-unmatched validation). The possible reason is that CPS is independent from IOP as the result of determined from IOP-matched scenario. That is, whether IOP matched or unmatched, CPS possibly reflects the true difference of corneal biomechanics between keratoconus and normal groups. Since the results were obtained from leave-one-out cross-validation, we believe that the compatibility of CPS is strongly supported.

In order to analyze the superiority, CPS was compared with aCBI [7] and DCR [8] in same IOP-matched and -unmatched validation. The results showed that CPS exhibited similar or better performance in each validation set. In our opinion, this is primarily caused by 4 reasons. First, CPS contains a new biomechanical parameter Eload [2]. It describes the energy loading during corneal deformation, which reflects the viscoelasticity of cornea and can essentially distinguish between keratoconus and normal cornea. Second, DRM1 is a known biomechanical parameter describing the ability of deformation, and its great impact on keratoconus screening has been proved in previous studies [6-8]. Third, in general, in a classification problem, the simpler the model, the better the robustness. CPS only has 3 parameters, less than aCBI or DCR, so it is likely to perform well on incoming new datasets. Finally, since CPS was chosen by the IOP-matched group, it may describe the true difference between the keratoconus and the normal corneal. Overall, the superiority of CPS has been further supported and validated by compared with other parameter sets.

It's worth noting that the main goal of our study was to validate that whether a parameter set determined in IOP-matched scenario can show the compatibility and superiority in both IOP-matched and -unmatched cases. However, it's not a trivial to evaluate its superiority without comparison with the already established parameter set aCBI and DCR. The results, which indicated CPS had the similar or better classification capacity than other parameter sets, can be expected to facilitate the further progress in keratoconus screening.

It's should be noted that we deliberately didn't match CCT because we didn't want to neglect the change of biomechanical stability as the result of corneal thinning as a part of the natural pathogenesis of keratoconus [8].

A potential limitation of this study is that the sample size in the IOP-matched validation was relatively small and therefore the results compared with IOP-unmatched validation may need to be interpreted cautiously.

In conclusion, our study demonstrated that, the parameter set (CPS), determined from IOP-matched scenario, is effective to distinguish keratoconus from normal corneas regardless of IOP-matched and -unmatched. More samples are warranted to elucidate its full usefulness in clinical practice, especially the early diagnosis of keratoconus.

**Table 1**. Parameters from the Corvis ST Software and Corresponding Definitions.
**Table 2**. Comparing the bIOP [23], CCT and Biomechanical Parameters in the Keratoconus and Normal corneas in IOP-matched Scenario (35 keratoconus and 35 normal eyes; pairwise matching for bIOP).
**Table 3.** Gain in Sensitivity and Specificity with Each Step of the Logistics Regression in bIOP-matched Scenario (35 keratoconus and 35 normal eyes; pairwise matching for bIOP) (best cut-off point = 0.5).
**Figure 1.** Receiver Operator Characteristic Curve for Each Step of the Forward Logistics Regression in IOP-matched Scenario (36 keratoconus and 36 normal eyes; pairwise matching for bIOP).
**Figure 2.** Receiver Operator Characteristic Curve for CPS, aCBI [7] and DCR [8], in IOP-matched (31 keratoconus and31normal eyes; pairwise matching for bIOP) and -unmatched (112 keratoconus and 91 normal eyes; not pair matching for bIOP) validation.
**Figure 3.** Accuracy, F1, sensitivity and specificity of CPS, aCBI [7] and DCR [8], in IOP-matched (31 keratoconus and31 normal eyes; pairwise matching for bIOP) and -unmatched (112 keratoconus and 91 normal eyes; not pair matching for bIOP) validation.